\documentclass[12pt]{article}
\def \1{\'{\i}}

\def \d{\displaystyle}

\def \p{Painlev\'e }

\def \&{&=&}
\def \t{\paragraph{$\bullet$}}
\def \nn{\nonumber}
\newcommand{\be}{\begin{equation}}
\newcommand{\ee}{\end{equation}}
\newcommand{\beq}{\begin{eqnarray}}
\newcommand{\eeq}{\end{eqnarray}} 
\newcommand{\ba}{\begin{array}}
\newcommand{\ea}{\end{array}}  

\begin{document}
\setcounter{equation}{0}
\setcounter{section}{0}
\title{\Large \bf Darboux Transformation and solutions for an equation in 2+1 dimensions}
\author{ \bf P.G.
Est\'evez\footnote{e-mail: pilar@sonia.usal.es}  \\ {\small
\bf Area de F\1sica Te\'orica}\\ {\small \bf Facultad de F\1sica}\\ {\small \bf Universidad de
Salamanca}\\  {\small \bf 37008 Salamanca. Spain}\\} 
\maketitle
\begin{abstract} 

 Painlev\'e analysis and the  singular manifold method are the tools used in this paper to
perform a complete study of an equation in 2+1 dimensions. This procedure has allowed us to obtain
the Lax pair, Darboux transformation and $\tau$ functions in such a way that a plethora of different
solutions with solitonic behavior  can be constructed iteratively

\end{abstract} \vspace*{0.3in}

{\bf PACS Numbers 02.30 and 03.40K}

\section{Introduction}
\setcounter{equation}{0}
Among the various approaches followed to study the behavior of non-linear partial differential
equations (PDEs),
\p analysis has proved to be one of the most   fruitful,  providing an algorithmic
procedure that affords a systematic way to deal with non-linear PDEs. Despite this,  it has
often 
been  used merely as a test of integrability while other methods, as Hirota's method or inverse
scattering, have been  used to obtain explicit solutions.

Our aim here  is to show, for an equation in 2+1, that an approach based on \p techniques,
such  as  the singular manifold method (SMM), can be successful in identifying many of the
properties of non-linear  PDEs (B\"acklund and Darboux transformations, $\tau$-functions, etc)
as well as  in constructing an iterative procedure to obtain multisolitonic solutions.

	The subject of our study is the 2+1 PDE  
\beq \nn 0\& V_y-(u\omega)_x\\ \nn 0\& \lambda u_t+u_{xx}-2uV\\
0\& \lambda \omega_t-\omega_{xx}+2\omega V \eeq 

The real version of this equation was obtained in \cite{Chakravarty}  as a reduction of
self-dual Yang-Mills equations  while the complex version appears in \cite{Maccari}.
The equation has the Painlev\'e property (PP) as it has been shown by Radha and Lakshmanan
\cite{Radha}  (real version) and Porsezian \cite{Porsezian} (complex version). The bilinear method 
was applied in \cite{Radha}  to obtain some soliton and dromion solutions.

For $\lambda =i$ and $\omega=u^*$, equation (1.1) is the expression proposed by Fokas in
\cite{Fokas}. This case contains the non-linear  Schr\"odinger equation when $x=y$ \cite{CE98}.

Recently, \cite{CE98} the author and her coworker have shown that there is a Miura
transformation between (1.1) and the generalized dispersive long wave equation \cite{Boiti},
\cite{AC}.

The plan of this paper is as follows: In Section II we shall   apply the
singular manifold method to equation (1.1). Section III, IV and V are devoted to showing how
the SMM allows us to determine algorithmically  the Lax pair as well as Darboux transformations
and
$\tau$-functions. In section VI, several solutions are  constructed explicitly. We close  with
a list of conclusions

\section{The Singular Manifold Method}
\setcounter{equation}{0}
The equation under study in this paper is the real version of (1.1), which reads:
\beq \nn 0\& m_y+u\omega\\ \nn 0\& u_t+u_{xx}+2um_x\\
0\& \omega_t-\omega_{xx}-2\omega m_x \eeq

where we have set $\lambda=1$ and $V=-m_x$.
\subsection{Leading term analysis}
To check the  the \p property \cite{PP} for (2.1), we require a generalized Laurent expansion  of
the fields  in terms of an arbitrary singularity manifold (depending on the initial data)
$\chi(x,y,t)=0$. This expansion should be of the form \cite{WTC}
\beq \nn u&=&\sum_{j=0}^{\infty}u_j\chi^{j-a}\\ 
\nn \omega&=&\sum_{j=0}^{\infty}\omega_j\chi^{j-b} \\ m&=&\sum_{j=0}^{\infty}m_j\chi^{j-c}\eeq
By substituting (2.2) in (2.1), we have for the leading terms:

\be a=b=c=1\qquad\qquad m_0=\chi_x,\qquad\qquad u_0\omega_0=\chi_x\chi_y\ee

Leading analysis  is able to determine the product of the dominant terms $u_0\omega_0$ but not
each one independently, which means that $u$ and
$\omega$ are not good fields in which to apply the singularity analysis because their dominant
behavior is not well defined. However,  for the field $m$, the leading term
$m_0$ is well defined. This  suggests  that the ``good field", from the point of view of the \p
analysis, is $m$. Accordingly, our first aim will be to write (2.1) as a partial differential 
equation only for
$m$.  It is not difficult to check (see appendix) that if we identify
\be m_t=n_x\ee
we can remove $u$ and $\omega$ from (2.1) to obtain the PDE:
\be
 0=m_y^2\left(n_{yt}-m_{xxxy}\right)+m_{xy}\left(n_y^2-m_{xy}^2\right)
+2m_y\left(m_{xy}m_{xxy}-n_yn_{xy}\right)-4m_y^3m_{xx}\ee
In \cite{CE98}, it has been shown that there is a Miura   transformation between (2.5) and the
generalized  long  dispersive wave equation \cite{Boiti}, \cite{AC}. This is why below we
shall be referring, in the next, to (2.5)  as MGLDW (modified generalized long  dispersive
wave equation). The  study of this equation for the field $m$  will be the subject of the rest
of this paper. Furthermore, $u$ and 
$\omega$  can  easily be obtained  from $m$ as:
\beq  u\&\sqrt{m_y}\quad e^{\displaystyle{\int{n_y\over 2m_y}dx}}\\
\nn \omega\&-\sqrt{m_y}\quad e^{-\displaystyle{\int{n_y\over 2m_y}dx}}\eeq
as we show in a detailed manner in the Appendix
 
\subsection{Truncated expansion. Auto-B\"acklund transformations}
As stated  above, the singularity manifold $\chi$ is an arbitrary function depending on the
initial data. The SMM requires us to restrict ourselves to the particular cases of the 
singularity manifold for which the expansion (2.2) truncates at  the constant level
\cite{Weiss}. In this  case the singularity manifold is not longer  an arbitrary function
because it is ``determined" by the condition of truncation.  We call it ``singular manifold"
and we shall use $\phi$ to refer to it. Thus, the truncation of (2.2) is: 
\be m'=m+{\phi_x\over \phi}\Longrightarrow
 n'=n+{\phi_t\over \phi}\ee
where both $m$ and $m'$ are solutions of (2.5). Accordingly,  truncation of the \p series
adopts the form of an 
auto-B\"acklund transformation between two solutions of (2.5).

\subsection*{Expression of the solutions in terms of the Singular  Manifold}

Substitution of (2.7) in (2.5) provides a polynomial in $\phi$. The way to proceed in the SMM
is to require that all the coefficients of this polynomial should be zero. The result should
be:

 a) The expression of $m$ in terms of  $\phi$. 

b) The equations to be satisfied by  $\phi$.

\medskip

For equation(2.5), the polynomial in  $\phi$ is rather complicated. We  used  MAPLE to handle
the calculation. This allows us to obtain the derivatives of  $m$ in terms of the singular
manifold. The result is: 
\beq  4m_x\& p_t-v_x-{v^2+w^2\over 2}\\
4n_y\& 2{(q_x+qv)_xp_y-(q_x+qv)p_{xy}\over q}+4{(p_y+qp_x)\over q}m_y \\ 4m_y\&
{p_y^2-(q_x+qv)^2\over q}
\eeq
where $p, q, w$ y $v$ are defined from the singular manifold as:
\beq  \nn v\&{\phi_{xx}\over \phi_x}\\  w\&{\phi_{t}\over
\phi_x}= p_x\\  \nn q\&{\phi_{y}\over \phi_x}\eeq
\subsection*{Singular manifold equations}
The equations to be satisfied by the singular manifold  $\phi$ are not difficult to obtain :
\t On one hand, there are some generic equations arising from the compatibility of the
definitions  (2.11). These are :
\beq \nn \phi_{xxt}&=\phi_{txx}&\Longrightarrow v_t=(w_x+vw)_x\\
\phi_{xxy}&=\phi_{yxx}&\Longrightarrow v_y=(q_x+vq)_x\\
 \nn \phi_{yt}&=\phi_{ty}&\Longrightarrow q_t=w_y+wq_x-qw_x  \eeq
\t Also, there is an  equation that is specific for (2.5) that can be determined by taking the
cross derivatives  in (2.8)-(2.10). This equation is:

\be p_{yt}=q_{xxx}+q(v_{xx}-vv_x)+p_xp_{xy}+\left({p_y^2-q_x^2\over q}\right)_x\ee
The set   (2.12)-(2.13) forms the singular manifold equations.
\section{Lax pair and SMM}
\setcounter{equation}{0}
It is  unnecessary to talk about the importance of determining the Lax pair of a
non-linear PDE. Nevertheless, in most   cases it is  determined  by inspection. We
shall see here that a non trivial advantage of  \p analysis is that it allows us to determine
the  Lax pair in an algorithmic way \cite{EG97}, \cite{ECG98}. 
 
\subsection{Dominant terms in singular manifold equations}
Returning to the singular manifold equations (2.12)-(2.13),  these can be considered to be
 a system of coupled non linear PDEs. We can therefore analyze their leading terms. This
requires us to set: 
\beq \nn w&\sim& w_0\chi^a\\
   v&\sim& v_0\chi^b \\ \nn q&\sim&q_0\chi^c\eeq
The balance of leading powers yields: 
\be a=b=-1\qquad\qquad c=0\ee
which means that only  $w$ and  $v$ have an expansion in negative powers of
$\chi$. Thus, the \p expansion is only pertinent for them but not for $q$. Moreover, the
leading analysis provides the leading coefficients
$w_0$ y $v_0$:
\beq \nn w_0&=&\pm \chi_x\\
v_0&=&\chi_x\eeq 
The $\pm$ sign of $w_0$ means that there are two possible \p expansions: The problem of systems
with two \p branches has been extensively discussed in  \cite{CMP},
\cite{EG93},
\cite{EG97} and  \cite{ECG98}. The suggestion of the author and coworker is that, for this class of
systems, it is necessary to consider both branches simultaneously by using two singular
manifolds;
 one for each branch.  
\subsection{Eigenfunctions and  the  singular manifold}
 With this idea in mind,  for the dominant terms of $w$ and  $v$ we can write: 
\beq  \nn v&=& {\displaystyle {\psi^+_x\over \psi^+}+{\psi^-_x\over \psi^-}}\\
w&=&{\displaystyle{\psi^+_x\over \psi^+}-{\psi^-_x\over \psi^-}} \eeq 
where we have used  $\psi^+$ for the singular manifold of the positive branch and 
$\psi^-$ for the negative one. As we will  seen later on, $\psi^+$ and $\psi^-$ will be the
eigenfunctions of the Lax pair and hereafter we will designate them as eigenfunctions.

\t Taking the derivatives of  (3.4) with respect to   $t$ and $y$ and using (2.12)  to integrate
them in 
$x$, we can write:
\beq \nn w_x+wv&=&{\displaystyle {\psi^+_t\over \psi^+}+{\psi^-_t\over \psi^-}}\\
p_t&=&{\displaystyle{\psi^+_t\over \psi^+}-{\psi^-_t\over \psi^-}}\eeq
and
\beq \nn q_x+qv&=&{\displaystyle {\psi^+_y\over \psi^+}+{\psi^-_y\over \psi^-}}\\
p_y&=&{\displaystyle{\psi^+_y\over \psi^+}-{\psi^-_y\over \psi^-}}\eeq

Expressions (3.4)-(3.6) allow us {\it to write the logarithmic derivatives of the eigenfunctions
$\psi^+$ and
$\psi^-$ in
terms of the singular manifold} as:

\beq  \alpha^+ &=\d{{\psi^+_x\over \psi^+}}=\d{{v+w\over 2}} \qquad\qquad\quad
 \alpha^- &=\d{{\psi^-_x\over\psi^-}={v-w\over 2}}\\ 
  \beta^+ &=\d{{\psi^+_y\over \psi^+}={q_x+qv+p_y\over 2}}\qquad
 \beta^- &=\d{{\psi^-_y\over\psi^-}={q_x+qv-p_y\over 2}}\\
  \gamma^+ &= \d{{\psi^+_t\over \psi^+}={w_x+wv+p_t\over 2}}\qquad
 \gamma^-&=\d{{\psi^-_t\over\psi^-}={w_x+wv-p_t\over 2}}\eeq
 $\alpha, \beta$ y $\gamma$ have been introduced with the single purpose of simplifying later
calculations
\t
Conversely, the determination of    $\phi$  from
$\psi^+$ y
$\psi^-$ is not  difficult taking into account (2.11),  which  allows us integrate (3.4) with
respect to 
$x$, which yields: 
\be  \phi_x=\psi^+\psi^-\ee
where the integration constant has been set at zero with no loss of generality (because the
singular manifold is defined except for a multiplicative constant). 
The $t$ derivative of $\phi$ can be obtained by combining  (2.11), (3.4) and  (3.10)
 to obtain:
\be \phi_t=\psi^-\psi_x^+-\psi^+\psi_x^-\ee
and, similarly, $\phi_y$ arises from (2.10), (2.11) and (3.6) as :
\be \phi_y=-{\psi_y^+\psi_y^-\over m_y}\ee

Equations (3.10), (3.11) and (3.12) allow us to construct $\phi$ from $\psi^+$ and $\psi^-$.
Accordingly,  {\it the total correspondence between singular manifolds and eigenfunctions is
explicitly constructed}.

\subsection{Linearization of the singular manifold equations: The Lax pair}

We return to equations (2.8)-(2.10). These equations are the expression of the seminal
solution $m$ in terms of the singular manifold. At the same time,  (3.7)-(3.12) relate the singular
manifold to the eigenfunctions. The question is now: How can we  express $m$ in terms of 
 $\psi^+$ and $\psi^-$?

\t As a previous step, it is easy to see that (2.10) can be combined with (3.8), yielding:
\be m_y=-{\beta^+\beta^-\over q}\Longrightarrow {\psi_y^+\psi_y^-\over \psi^+\psi^-}=-qm_y\ee
which shows the coupling between $\psi^+$ and
$\psi^-$.

\t Let us return to (2.8).  To write this in terms of  the eigenfunctions,
we need to substitute   $v$ and
$p_t$  from (3.7) and (3.9)
$$ 4m_x= 2\gamma^+-2\alpha_x^+-2(\alpha^+)^2,\qquad {\rm or} \qquad 4m_x=
-2\gamma^--2\alpha_x^--2(\alpha^-)^2$$ Now, by substituting $\alpha$ and $\gamma$:

\be 0=\psi^+_t-\psi^+_{xx}-2m_x\psi^+,\qquad\qquad {\rm or} \qquad\qquad 
0=\psi^-_t+\psi^-_{xx}+2m_x\psi^-\ee and this can be considered as  the temporal part of the
Lax pair.

\t Finally, by combining it with (3.7)-(3.8) (2.9) can be written as:
\be qn_y=
\left[\beta^+\beta_x^--\beta^-\beta 
_x^+\right]
+m_y\left[(\beta^+-\beta^-)+q(\alpha^+-\alpha^-)\right]\ee
If we use  $\beta^+\beta^-=\d{-qm_y}$ and
$\alpha^++\alpha^-=v$ to remove from (3.15) ($\beta^-$, $\alpha^-$) or ($\beta^+$, $\alpha^+$).
$$ n_y=-m_{xy}+2m_y\left({\beta_x^+\over \beta^+}+\alpha^++{m_y\over \beta^+}\right),\qquad
n_y=m_{xy}-2m_y\left({\beta_x^-\over \beta^-}+\alpha^-+{m_y\over \beta^+}\right)$$ or
\be (n_y+m_{xy})\psi_y^+=2m_y(\psi_{xy}^++m_y\psi^+),\qquad\qquad
(-n_y+m_{xy})\psi_y^-=2m_y(\psi_{xy}^-+m_y\psi^-)\ee and this can  be considered the spatial
part of the Lax pair.

Thus, the SMM allows us to define two eigenfunctions,  $\psi^+$ and $\psi^-$, such
that {\it the expression of the truncated solutions  in terms of these eigenfunctions is precisely the
Lax pair} (3.15)-(3.16).

\section{Darboux Transformations}
\setcounter{equation}{0}
This section will be devoted to determining an algorithmic procedure for constructing solutions.
 
\t We shall summarize the results obtained in the previous section: Let  $m$  be a solution of
(2.5), and 
$\phi_1$ a singular manifold for it. This singular manifold can be constructed by means of two
eigenfunctions $\psi_1^+$ and $\psi_1^-$ through:
\beq
\nn \phi_{1x}\& \psi_1^+\psi_1^-\\
\nn m_y\phi_{1y}\&-\psi_{1y}^+\psi_{1y}^-\\
 \phi_{1t}\&  \psi_1^-\psi_{1x}^+-\psi_1^+\psi_{1x}^-\eeq
where $\psi_1^+$ and $\psi_1^-$ satisfy the Lax pairs:

\beq &0= \psi_{1t}^+-\psi_{1xx}^+-2m_x\psi_1^+\quad\quad\qquad
\qquad\qquad &0=\psi_{1t}^-+\psi_{1xx}^-+2m_x\psi_1^-\\ \nn&0=
2m_y\psi_{1xy}^+-(m_{xy}+n_y)\psi_{1y}^++2m_y^2\psi_1^+
\qquad &0= 2m_y \psi_{1xy}^- -(m_{xy}-n_y)\psi_{1y}^-+2m_y^2\psi_1^-\eeq
\t According to (2.7), the singular manifold $\phi_1$ allows us  to define a new solution
$m'$
\be m'=m+{\phi_{1x}\over \phi_1}\Longrightarrow n'=n+{\phi_{1t}\over \phi_1}\ee
whose Lax pairs will be

\beq  &0=\psi^{'+}_{t}-\psi^{'+}_{xx}-2m'_x\psi^{'+}\quad\qquad\qquad  &0=
\psi^{'-}_{t}+\psi^{'-}_{xx}+2m'_x\psi^{'-}\\ \nn &0=
2m'_y\psi^{'+}_{xy}-(m'_{xy}+n'_y)\psi^{'+}_{y}+2m_y^{'2}\psi^{'+}\qquad &0=2m'_y
\psi^{'-}_{xy}-(m'_{xy}-n'_y)\psi^{'-}_{y}+2m_y^{'2}\psi^{'-}\eeq

and $\psi^{'+}$ and
$\psi^{'-}$ can be used to construct, for $m'$, a singular manifold $\phi'$ defined as:
\beq
\nn \phi'_{x}\& \psi^{'+}\psi^{'-}\\
\nn m'_y\phi'_{y}\&-\psi_{y}^{'+}\psi^{'-}_{y}\\
 \phi'_{t}\&  \psi^{'-}\psi^{'+}_{x}-\psi^{'+}\psi^{'-}_{x}\eeq

\subsection*{Truncated expansion in the Lax pair}

A Lax pair such as (4.4)  is usually considered  to be a linear system  for $\psi'$, where $m'$ is
the potential and hence the inverse scattering method can be applied.

A different interpretation  \cite{KS},  \cite{EG97} of (4.4)  is to consider it as  a coupled 
``non linear" system of PDEs for the fields $m', n', \psi^{'+}, \psi^{'-}$. In this case, the
singular manifold method can be applied to the Lax pair itself and the  truncated expansion
(4.3) for $m$ and $n$ should be combined with a similar expansion for $\psi^{'+}$ and $\psi^{'-}$.
In fact, this expansion could be written as: 
$$ \psi^{'+}=\psi^+_2+{\psi^+_0\over \phi_1}\qquad \psi^{'+}=\psi^-_2+{\psi^-_0\over \phi_1}$$
where $\psi^+_0, \psi^-_0$ are the dominant terms. It is useful. For later calculations, it is
useful to set 
\break 
$\psi^+_0=-\psi^+_1\Omega^+$, and 
$\psi^-_0=-\psi^-_1\Omega^-$. Therefore:
\be \nn \psi^{'+}=\psi^+_2-\psi^+_1{\Omega^+\over \phi_1}\qquad
\psi^{'-}=\psi^-_2-\psi^-_1{\Omega^-\over\phi_1}\ee

Substitution of the truncated expansions (4.3) and (4.6) in the Lax pairs (4.4)   provides the
following results  (we used  MAPLE for the calculations):
\t 1) $\psi^+_2$ and  $\psi^-_2$ are eigenfunctions for  $m$. Consequently, they satisfy Lax
pairs such as:

\beq &0= \psi_{2t}^+-\psi_{2xx}^+-2m_x\psi_2^+\quad\quad\qquad
\qquad\qquad &0=\psi_{2t}^-+\psi_{2xx}^-+2m_x\psi_2^-\\ \nn&0=
2m_y\psi_{2xy}^+-(m_{xy}+n_y)\psi_{2y}^++2m_y^2\psi_2^+
\qquad &0= 2m_y \psi_{2xy}^- -(m_{xy}-n_y)\psi_{2y}^-+2m_y^2\psi_2^-\eeq

\t 2) $\Omega^+$ and $\Omega^-$ are related to the eigenfunctions in the following way:
\beq
\nn \Omega^-_x\&\psi^+_1\psi^-_2\\
\nn m_y \Omega^-_y\&-\psi^+_{1y}\psi^-_{2y}\\
\Omega^-_t\&\psi^+_{1x}\psi^-_2-\psi^+_1\psi^-_{2x}\eeq
\beq
\nn \Omega^+_x\&\psi^+_2\psi^-_1\\
\nn m_y \Omega^+_y\&-\psi^+_{2y}\psi^-_{1y}\\
\Omega^+_t\&\psi^+_{2x}\psi^-_1-\psi^+_2\psi^-_{1x}\eeq

\t To summarize: Two pairs of eigenfunctions 
$(\psi_1^+,\psi_1^-)$,
$(\psi_2^+,\psi_2^-)$ for a solution  $(m,n)$ are sufficient  to construct the following
transformation:,
\beq
\nn m'\&m+{\phi_{1x}\over \phi_1}\\
\nn n'\&n+{\phi_{1t}\over \phi_1}\\
\nn \psi^{'+}\&\psi^+_2-\psi^+_1{\Omega^+\over \phi_1}\\
\psi^{'-}\&\psi^-_2-\psi^-_1{\Omega^-\over\phi_1}\eeq
where  $\phi_1$, $\Omega^+$ and $\Omega^-$ are related to the eigenfunction through (4.1), (4.8) and
(4.9).

Equation (4.10) is a transformation of potentials and eigenfunctions that leaves invariant the
Lax pairs. It should therefore  be considered  a {\it Darboux transformation} \cite{MS}. 

\section{Iteration of the singular manifold: $\tau$-functions}
\setcounter{equation}{0}

A well known method for obtaining  multisolitonic solutions of  PDEs is the bilinear Hirota
method. Indeed, some solutions of (2.1) have been identified with this method
\cite{Radha}. Let us address ourselves to the task of  establishing, by explicit construction, 
the relationship between the singular manifold and the $\tau$-functions of  Hirota's
method. 

\t  Equation (4.5) could be considered as a
non linear system among $\phi'$, $\psi^{'+}$ and  $\psi^{'-}$. For this system we can use   the
same criterion  used in the previous section. It requires that the expansion 
\beq \nn \psi^{'+}\&\psi^+_2-\psi^+_1{\Omega^+\over \phi_1}\\
\psi^{'-}\&\psi^-_2-\psi^-_1{\Omega^-\over\phi_1}\eeq
for $\psi^{'+}$ and  $\psi^{'-}$ should be combined with a truncated expansion for $\phi'$.

\be \phi'=\phi_2+{\Delta\over \phi_1}\ee
It is not difficult to prove that the  substitution of (5.1)-(5.2) in (4.5) gives:
\be \Delta=-\Omega^+\Omega^-\ee
while  $\phi_2$ is the singular manifold for $m$ related to  $\psi_2^+$ and $\psi_2^-$, which means:
\beq
\nn \phi_{2x}\& \psi_2^+\psi_2^-\\
\nn m_y\phi_{2y}\&-\psi_{2y}^+\psi_{2y}^-\\
 \phi_{2t}\&  \psi_2^-\psi_{2x}^+-\psi_2^+\psi_{2x}^-\eeq
\t As far as (5.2) defines a singular manifold for $m'$,   it can be used to obtain a new solution:

\be m''=m'+{\phi'_x\over \phi'} \qquad n''=n'+{\phi'_t\over \phi'}\ee
which, combined with  (4.3 ), is: 
\be m''=m+{\tau_x\over \tau} \qquad n''=n+{\tau_t\over \tau}\ee
where
\be \tau=\phi'\phi_1=\phi_1\phi_2-\Omega^+\Omega^-\ee
In the previous section we have shown that $\phi_1$, $\phi_2$, $\Omega^+$, $\Omega^-$ are obtained
from the eigenfunctions
$(\psi_1^+,\psi_1^-)$, $(\psi_2^+,\psi_2^-)$. Therefore {\it Equation (5.7) affords  the
relationship between $\tau$-functions, on one hand,  and singular manifolds,} on the other hand.

\section{Solutions}
\setcounter{equation}{0}

From the previous results we can derive an iterative  procedure to construct solutions. It can be
summarized as follows:

1) Starting with a seminal solution $m$, solve the Lax pairs (4.2) and (4.8) to obtain
$\psi_1^+$, $\psi_1^-$, $\psi_2^+$, $\psi_2^-$.

2) Perform the integration of (4.1), (4.8), (4.9) and (5.4) to get $\phi_1$, $\Omega^+$,
$\Omega^-$ and $\phi_2$: Use (5.7) to construct 
$\tau$

3) Use (4.3) to obtain the solution $m'$ for the first iteration and (5.5) for the second one
$m''$.

The easiest way to obtain explicit solutions is to apply the above explained procedure, starting
with  a trivial seminal solution. We shall use as seminal
solutions $m=m_0y$ and $m=0$. From the dependence on $y$ of  (4.1), (4.9), (4.10) and (5.4) it is
 that the behavior is totally different, depending  on whether $m_y$  is zero or
not and giving rise to line-soliton or dromion behavior, respectively. 

\subsection{Line solitons $m=\omega_0 y$}
\t The easiest solutions of (4.2) and (4.7) are:
\beq &\nn \psi_1^+=\exp \left[a_1^+x-{\omega_0\over a_1^+}y+a_1^{+2}t\right] \qquad
&\psi_2^+=\exp \left[a_2^+x-{\omega_0\over a_2^+}y+a_2^{+2}t\right]\\ 
&\psi_1^-=\exp \left[a_1^-x-{\omega_0\over a_1^-}y-a_1^{-2}t\right]\qquad &\psi_2^-=\exp
\left[a_2^-x-{\omega_0\over a_2^-}y-a_2^{-2}t\right]\eeq 
where  $a_1^+, a_1^-, a_2^+$ y $a_2^-$ are arbitrary constants. 
\t Integration of (4.1), (4.8), (4.9) and (5.4) affords:
\beq \nn &\phi_1={1\over a_1^++a_1^-}(b_1+\psi_1^+\psi_1^-)\qquad
&\Omega^+={1\over a_2^++a_1^-}(c^++\psi_1^+\psi_2^-)
\\
&\phi_2={1\over
a_2^++a_2^-}(b_2+\psi_2^+\psi_2^-)\qquad &\Omega^-={1\over a_1^++a_2^-}(c^-+\psi_2^+\psi_1^-)\eeq
where  $b_1, b_2, c^+$ y $c^-$ are arbitrary constants
\t
The first iteration provides the solution (figure 1):
\be m'_y=\omega_0+\partial_{xy}[\ln \phi_1]\ee
and the second (figure 2):
\be m''_y=\omega_0+\partial_{xy}[\ln \tau]\ee
where
\be \phi_1={b_1\over a_1^++a_1^-}(1+F_1)\ee
\be \tau=\phi_1\phi_2-\Omega^+\Omega^-={b_1b_2\over
(a_1^++a_1^-)(a_2^++a_2^-)}\left[1+F_1+F_2+AF_1F_2\right]\ee
and

\be F_i(x,y,t)=\exp \left[(a_i^++a_i^-)\left\{x-{\omega_0\over
a_i^+a_i^-}y+(a_i^+-a_i^-)t\right\}+\varphi_i\right]\ee

\be A={(a_2^+-a_1^+)(a_2^--a_1^-)\over (a_2^++a_1^-)(a_2^-+a_1^+)}\ee
and $b_i$ has been redefined as:
$b_i=e^{-\varphi_i}$

\paragraph{Particular case:}When  $a_2^+=a_1^+$, or $a_2^-=a_1^-$, $A=0$ and this is said to be 
{\it resonant state}
\cite{EL95}
\subsection{Dromions $m=0$}
In this case (4.1), (4.8), (4.9) and (5.4) require that
$$\psi_{1y}^+\psi_{1y}^-=\psi_{2y}^+\psi_{2y}^-=\psi_{1y}^+\psi_{2y}^-=\psi_{2y}^+\psi_{1y}^-=0$$
Therefore it is compulsory that   $\psi_{1y}^-=\psi_{2y}^-=0$, or  $\psi_{1y}^+=\psi_{2y}^+=0$
\t If we choose the possibility $\psi_{1y}^-=\psi_{2y}^-=0$, then simple solutions of (4.2) and
(4.7) are:
\beq\nn &\psi_1^-= e^{\displaystyle{a_1^-x-a_1^{-2}t}}\quad
&\psi_1^+=\left(e^{\displaystyle{a_1^+x+a_1^{+2}t}}\right)E_1(y)=Q_1^+(x,t)E_1(y)\\
&\quad \psi_2^-= e^{\displaystyle{a_2^-x-a_2^{-2}t}}\qquad
&\psi_2^+= \left(e^{\displaystyle{a_2^+x+a_2^{+2}t}}\right)E_2(y)=Q_2^+(x,t)E_2(y)\eeq
where  $a_1^+, a_1^-, a_2^+, a_2^-$ are arbitrary constants  while $E_i$ are arbitrary functions of
$y$

\t We can now  perform now the integration of (4.1), (4.8), (4.9) and (5.4) to obtain:
\beq \nn &\phi_1={\displaystyle{1\over a_1^++a_1^-}}\left(E_1Q_1^+\psi_1^-+M_1(y)\right)\qquad
&\Omega^+={1\over a_2^++a_1^-}\left(E_2\psi_1^-Q_2^++N^+(y)\right)\\
&\phi_2={\displaystyle{1\over a_2^++a_2^-}}\left(E_2\psi_2^-Q_2^++M_2(y)\right)\qquad
&\Omega^-={1\over a_1^++a_2^-}\left(E_1\psi_2^-Q_1^++N^-(y)\right) \eeq 
$N^+$, $N^-$ and $M_i^+$ are arbitrary functions of $y$. The arbitrariness of the six functions
$E_i$, $M_i$, $N^j$  and the four constants $a_i^+$, $a_i^-$ implies that there are many 
particular cases. We  list some of them: 

\subsubsection{1+1 dromions:} These  can be obtained by choosing
\be
E_i(y)=1+b_ie^{c_iy}\qquad\qquad M_i(y)=1+e^{c_iy}\qquad\qquad N^*=N^-=0\ee
where $b_i$ and $c_i$ are arbitrary constants.
\t The first and second iteration yield:
\be m'_y=\partial_{xy}[\ln \phi_1]\ee \be m''_y=\partial_{xy}[\ln \tau]\ee
where
\be \phi_1={1\over a_1^++a_1^-}(M_1+E_1F_1)\ee
\be \tau=\phi_1\phi_2-\Omega^+\Omega^-={1\over
(a_1^++a_1^-)(a_2^++a_2^-)}\left[M_1M_2+M_1E_2F_2+M_2E_1F_1+AE_1E_2F_1F_2\right]\ee
and

\be F_i(x,t)=\exp \left[(a_i^++a_i^-)\left\{x+(a_i^+-a_i^-)t\right\}\right]\ee

\be A={(a_2^+-a_1^+)(a_2^--a_1^-)\over (a_2^++a_1^-)(a_2^-+a_1^+)}\ee

The behavior of (6.12) and (6.13) are represented in Figure 3 and Figure 4 respectively.

\subsubsection{1+n dromion:} Dromions with several jumps in the $y$ direction can be obtained by
choosing
$$
E_i=1+\sum_{j=1}^n b_{ij}e^{{\displaystyle c_{ij}y}}\qquad\qquad M_i=1+\sum_{j=1}^n
e^{{\displaystyle c_{ij}y}}$$
The first iteration 
$$ m'_y=\partial_{xy}(\ln \phi_1)$$describes a structure with n jumps located along the
$y$-direction, moving in the $x$-direction with velocity $a_1^+-a_1^-$

Figure 5  represents one of these structures with $n=2$ and $c_{11}>0$, $c_{12}<0$ 

Figure 6  corresponds to  $n=3$ and $c_{11}>0$, $c_{12}>0$, $c_{13}>0$.  

The solution that we have obtained in this section generalizes the solutions found in \cite{Radha}
by means of the bilinear method

\section*{Conclusions}
\t A system of non-linear PDEs proposed by different authors as one of the simplest equations in
2+1 dimensions is studied from the point of view of the Painlev\'e property. The dominant behavior
indicates  the best field to use Painlev\'e analysis. On this basis, we rewrite the system
as a PDE (2.5) with only  one field. This equation \cite{CE98} can be considered as the modified
version of the generalized long dispersive wave equation \cite {Boiti}. This is why we have
call it MGLDW (modified generalized long dispersive wave equation)

\t The singular manifold method was applied to MGLDW in section II. The singular manifold
equations, as well as the expression of the seminal field in terms of the singular manifold were
obtained.

\t In section III, we  linearized the singular manifold equations to obtain the Lax pair. The
relation between the singular manifold and the eigenfunctions of the Lax pair is
constructed explicitly.

\t In section IV the Lax pair was considered as a system of non linear coupled PDE. We  
applied the singular manifold method to the Lax pair itself. The bonus  is  the construction
of Darboux transformations for MGLDW. Its transformations allow us to determine an iterative
method for obtaining solutions. The relation between this method and the Hirota $\tau$-functions
is  shown in section V

\t Finally section VI is devoted to the construction of solitonic solutions of MGLDW. A
rich collection of solutions with different  solitonic behavior appear depending on the seminal
solution that we have chosen

\t We believe that the equation  analyzed in depth in this paper is a good example of how to obtain
 maximum  information about the equation  using   Painlev\'e
analysis and the singular manifold method as the only tools.

\paragraph{ACKNOWLEDGEMENTS} This research has been supported in part by DGICYT  under contract
PB95-0947. I would like to thank Professor J. M. Cerver\'o for encouragement and a careful
reading of the manuscript
\appendix
\section*{Appendix}

We first attempt to write:

$$0= u_t+u_{xx}+2um_x\eqno(A.1)$$
 $$ 0= \omega_t-\omega_{xx}-2\omega m_x\eqno(A.2)$$
$$0=u\omega+m_y \eqno(A.3)$$
as an equation for only one field

Taking  $u$ out of  (A.3) and substituting it in (A.1), we obtain:
$$ 0={m_{yt}+m_{xxy}\over \omega}-2m_{xy}{\omega_x\over \omega^2}+m_y\left({2m_x\over
\omega}-{\omega_t\over \omega^2}-{\omega_{xx}\over \omega^2}-2{\omega_x^2\over
\omega^3}\right)\eqno(A.4)$$
 
\noindent Using (A.2) in (A.4), we also obtain:
$$ 0=m_{xxy}+m_{yt}-\left (2m_y{\omega_x\over \omega}\right)_x\eqno(A.5)$$
which can easily be integrated by setting $m_t=n_x$, which yields:
 
$$ {\omega_x\over \omega}={m_{xy}+n_y\over
2m_y}\eqno(A.6)$$ Substituting (A.6) in (A.2), we obtain:
$${\omega_t\over \omega}=2m_x+{m_{xxy}+n_{xy}\over 2m_y}-{m_{xy}^2-n_y^2\over 4m_y^2}\eqno(A.7)$$
Next, we calculate the  identity  $\displaystyle{\left({\omega_t\over
\omega}\right)_x=\left({\omega_x\over \omega}\right)_t}$ using (A.6) and (A.7), and finally we
obtain for $m$ the equation:
$$ 0= m_t-n_x\eqno(A.8)$$
$$0=m_y^2\left(n_{yt}-m_{xxxy}\right)+m_{xy}\left(n_y^2-m_{xy}^2\right)\\
+2m_y\left(m_{xy}m_{xxy}-n_yn_{xy}\right)-4m_y^3m_{xx}\eqno(A.8)$$
The integration of (A.6) is:
$$ u=\sqrt{m_y}\quad e^{\displaystyle{\int{n_y\over 2m_y}dx}}\eqno(A.9)$$
And from  (3.4)we finally obtain:
$$ \omega=-\sqrt{m_y}\quad e^{\displaystyle{\int-{n_y\over 2m_y}dx}}\eqno(A.10)$$

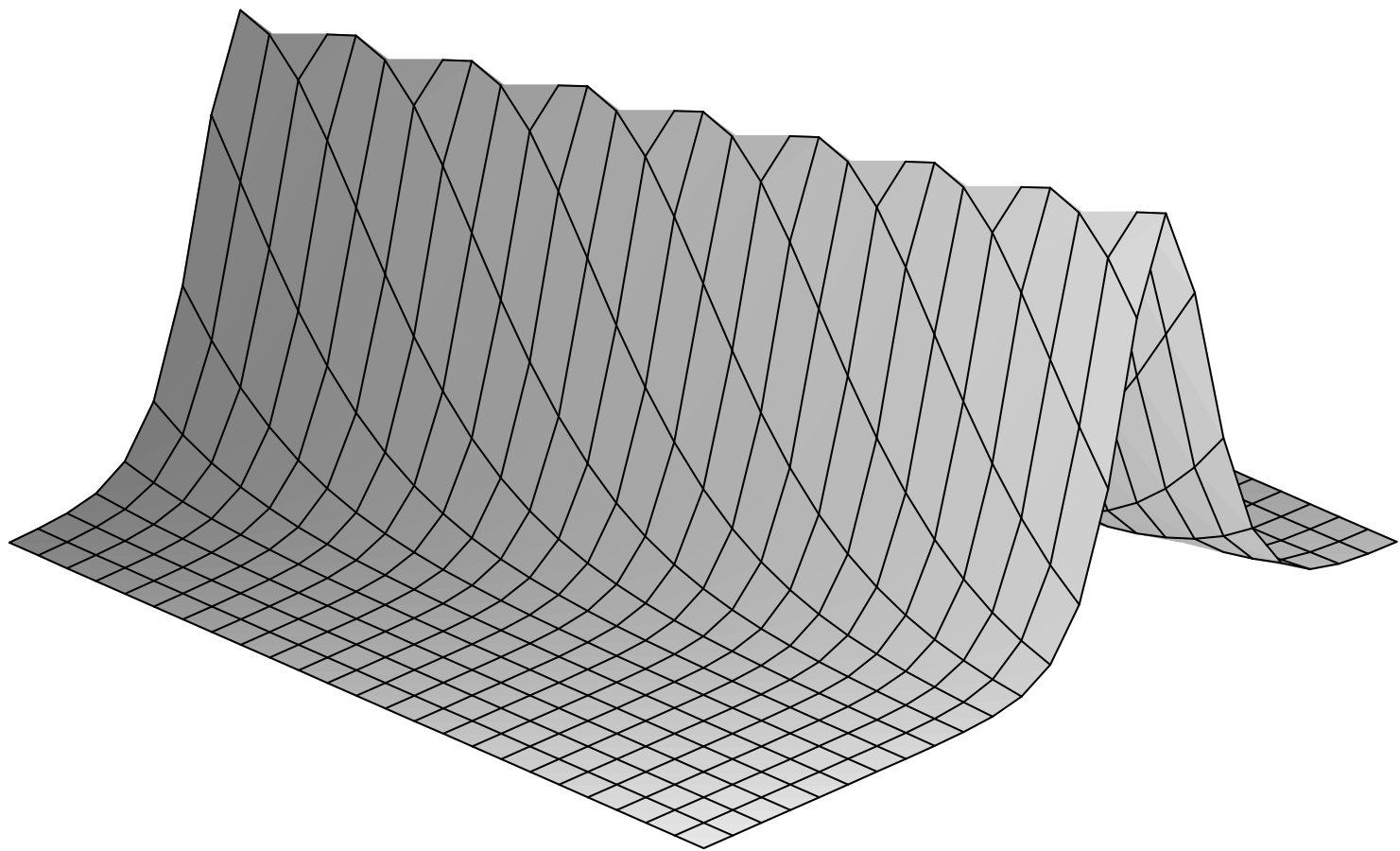
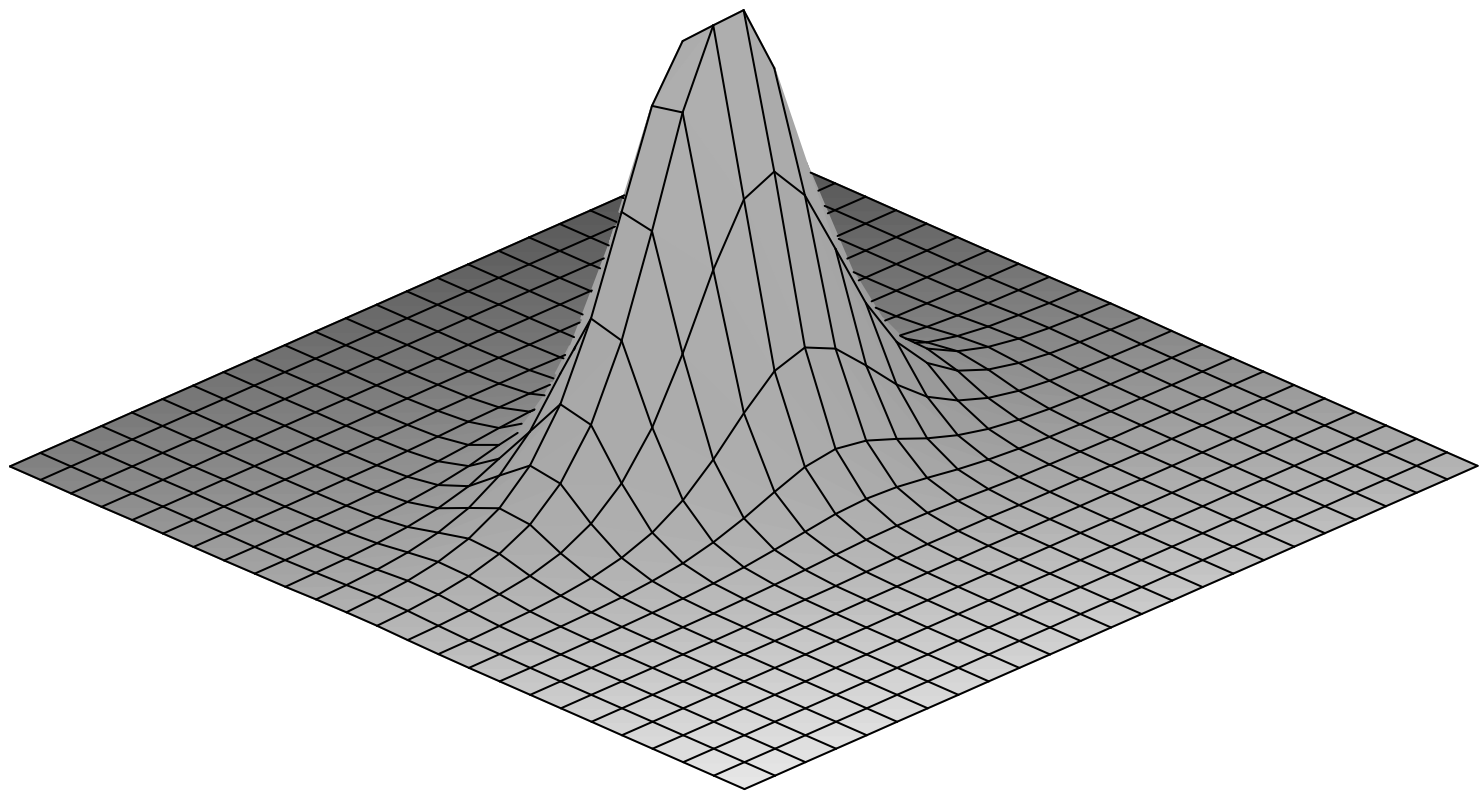
\end{document}